\documentclass[conference]{IEEEtran}
\usepackage{times}
\usepackage{xcolor}
\usepackage{graphicx}
\usepackage{booktabs}
\usepackage{afterpage}

\usepackage[numbers]{natbib}
\usepackage{multicol}
\usepackage[bookmarks=true]{hyperref}

\begin{document}

\title{Robot Character Generation and Adaptive Human-Robot Interaction with Personality Shaping}

% \author{Author Names Omitted for Anonymous Review. Paper-ID [766]}

\author{
    \authorblockN{Cheng Tang, Chao Tang, Steven Gong, Thomas M. Kwok, Yue Hu} \\
    \authorblockA{University of Waterloo, Waterloo, ON, Canada} \\
    Emails: \{cheng.tang, chao.tang, steven.gong, thomasm.kwok, yue.hu\}@uwaterloo.ca
}

\maketitle

\begin{abstract}

We present a novel framework for designing emotionally agile robots with dynamic personalities and memory-based learning, with the aim of performing adaptive and non-deterministic interactions with humans while conforming to shared social understanding. While existing work has largely focused on emotion recognition and static response systems, many approaches rely on sentiment analysis and action mapping frameworks that are pre-defined with limited dimensionality and fixed configurations, lacking the flexibility of dynamic personality traits and memory-enabled adaptation. Other systems are often restricted to limited modes of expression, such as verbal, facial, or gestural responses, and fail to develop a causal relationship between human behavior and the robot's proactive physical actions, resulting in constrained adaptability and reduced responsiveness in complex, dynamic interactions. Our methodology integrates the Big Five Personality Traits, Appraisal Theory, and abstracted memory layers through Large Language Models (LLMs). The LLM generates a parameterized robot personality based on the Big Five, processes human language and sentiments, evaluates human behavior using Appraisal Theory, and generates emotions and selects appropriate actions adapted by historical context over time. We validated the framework by testing three robots with distinct personalities in identical background contexts and found that personality, appraisal, and memory significantly influence the quality and adaptability of human-robot interactions. The significance of the individual components, as well as their integration, was further validated through ablation tests. We conclude that this system enables robots to engage in meaningful and personalized interactions with human users, transcending their role as mere tools. It holds significant potential for applications in domains such as pet robots, assistive robots, educational robots, and collaborative functional robots, where cultivating tailored relationships and enriching user experiences are essential.

\end{abstract}

\IEEEpeerreviewmaketitle

\section{Introduction}
Although robots are currently treated as functional tools in the industry \cite{manufacturing2015mobile, manufacturing2016skill, manufacturing2019collab, agriculture2021robots, agriculture2018robots, agriculture2021overview, warehouse2016growth}, the need for seamless and intuitive human-robot interaction will become necessary as robots increasingly integrate into industrial settings and interact more closely with humans. Furthermore, as robots are increasingly integrated into human-populated environments and workplaces, the importance of authentic interactions and meaningful relationships has been stressed to enhance user-interaction quality, including acceptance, satisfaction, and work efficiency \cite{customer-service2023, gorur2023fabric, anticipatory2007efficiency, pet2021emotion}. To establish meaningful interactions and authentic relationships, literature \cite{bui2019robot, emotion2022review, wellbeing2022review, abdollahi2022artificial} suggested three key aspects: (1) distinct robot personalities, (2) adaptive behavior based on past interactions, and (3) customizable emotional agility.

Distinct robot personalities can increase users' acceptance. For instance, research has shown that lifelike robot interactions in customer service with higher perceived values can enhance relationship quality, which leads to greater user satisfaction \cite{customer-service2023}. 
Adaptive behavior based on past interactions improves working efficiency with human users based on experiences. In collaborative settings, robots integrate more seamlessly as cohesive team members alongside human counterparts by intuitively adapting to user preferences and behaviors, thereby reducing communication overhead and improving collaboration efficiency \cite{gorur2023fabric, anticipatory2007efficiency}. 
Customizable emotional agility establishes trustworthy and profound relationships with users. In companionship-focused applications such as pet robots, emotional intelligence and personalized responses enhance user experiences while providing therapeutic benefits \cite{abdollahi2022artificial, pet2021emotion, pet2022theraputic, wellbeing2022review}.

Traditional approaches to affective robots, including personality modeling, sentiment analysis and response systems, and memory systems, rely heavily on predefined, domain-specific methods, limiting their adaptability and dimensionality in complex, dynamic scenarios. Sentiment analysis struggles with nuanced and ambiguous cues, while memory systems such as episodic and associative models fail to integrate contextual information or synthesize knowledge across interactions, reducing their ability to generalize and adapt. Although LLM-powered systems show promise in addressing these limitations by enabling context-aware, coherent, and evolving interactions, their application in robotics, particularly in handling physical actions and real-world adaptability, remains underexplored. By encompassing all three components, the system can capture, interpolate, and embody their interconnected relationships, enabling more lifelike and natural human-robot interactions.

Therefore, this paper aims to contribute towards the development of emotionally intelligent and adaptive robots that can engage in meaningful and personalized human-robot interactions. Inspired by psychological frameworks, it leverages LLMs to integrate the following three core components:
\begin{itemize}
    \item Big Five Personality Traits \cite{fivefactor1987validation, bigfive1992background} – Used for parameterizing robot personality.
    \item Appraisal Theory \cite{smith1990emotion, roseman2001appraisal, ellsworth2003appraisal} – Applied to evaluate human behavior.
    \item Memory \cite{park2023generative} – Enables adaptive mental and behavioral adjustments based on past interactions.
\end{itemize}
This framework allows the robot to generate emotions influenced by its personality and previous experiences, leading to contextually appropriate and dynamic responses to human input during interactions.

The rest of the paper is organized as follows: Section II discusses existing literature related to robot personality, emotional intelligence, and adaptation. Sections III describes the key novelty of our proposed system. Section IV introduces our proposed system. Section V presents experimental results. Sections VI and VII discusses the findings and limitations, and Section VIII concludes the paper.

% methodology

\section{Related Work}

\subsection{Enhancing Robot Lifelikeness with Personality}%Endowing Robots with Personality for Enhanced Lifelikeness}

Designing robots with personality and memory has been shown to significantly influence user modeling in human-robot interactions. This suggests that a robot’s personality traits can influence how it remembers and interprets user interactions, thereby enhancing personalization and making future interactions more natural \cite{personality2024adaptability}. The Big Five traits is the most widely used multidimensional framework for Personality on Robots (POR) \cite{personality2020review}. They are a widely recognized framework in psychology that categorizes human personality into five core dimensions: Openness, Conscientiousness, Extraversion, Agreeableness, and Neuroticism \cite{fivefactor1987validation, bigfive1992background}. We can effectively define robot's behavior and personality based on these trait. This demonstrates significant potential for authentic and lifelike interaction with human users.

Traditionally, robot personality has been expressed using explicit methods such as visual appearance, language, vocal features, movement, facial expressions, haptics, interaction patterns, and proxemics \cite{personality2020review}. Previous work \cite{personality2016behavior} highlighted that interaction authenticity and user trust can be enhanced by embedding personality traits into robots’ decision-making processes and ensuring consistent and predictable behaviors. Recent advancements have explored POR in the context of large language models (LLMs). For example, \cite{nardelli2024personality} introduces a framework that integrates personality traits to enhance emotional intelligence, enabling agents to exhibit consistent, contextually appropriate responses. This shows humans can accurately perceive and distinguish the robot’s personality. Collectively, these works have explored and validated the implementation and effects of incorporating robot personality, paving the way for integrating this with other domains, such as affective computing and adaptive systems, to achieve more natural human-robot interactions. 

\subsection{Developing Emotional Intelligence and Agility in Robots}

In addition to personality, emotional intelligence is another key component in enhancing human-robot interaction and establishment of a genuine and deep human-robot relationship \cite{stock2022survey, spezialetti2020emotion, emotion2022review, wellbeing2022review}. Many existing works attempted to make robots affectionate for companionship applications \cite{abdollahi2022artificial, pet2021emotion} and beyond \cite{stock2022survey, wang2023emotional, gorur2023fabric}. These applications typically encompass emotion recognition, empathetic response generation, and adaptive interaction, aiming at providing emotional support, fostering a sense of companionship, and enhancing user well-being \cite{abdollahi2022artificial, pet2021emotion}. Demonstrated by \cite{gorur2023fabric}, emotional agility allows collaborative robots to dynamically adapt behavior based on emotional context to offer more intuitive, cooperative, and efficient collaboration in shared workspaces.

The work in \cite{emotion2022review} reviews the implementation of emotional intelligence on robots, classifying techniques into 3 categories: sensing, computing, and acting. Sensing refers to recognizing human emotion using inputs including facial expression, gesture, and voice, and is commonly achieved with visual and acoustic feature extraction or deep learning techniques \cite{emotion2022review}. Computing and acting aspects of emotional intelligence refer to the processes by which the agents assess, understand, and respond to emotional stimuli to guide their behavior and decision-making adaptively \cite{emotion2022review}. Several works were proposed to address the development of computational frameworks and models for enabling agents to exhibit emotional intelligence \cite{jain2015emia, taverner2022genia3, ghafurian2021improving, wang2023emotional, zhao2024both}. The study \cite{jain2015emia} integrates appraisal theories of emotions and fuzzy logic to enable intelligent agents to elicit, regulate, and transition emotions continuously. By incorporating both regular affective and empathic appraisal processes, \cite{taverner2022genia3} proposes a framework that allows agents to select plans based on analyzed events, their own affective states, and personalities, while distinguishing between self and others. The work in  \cite{ghafurian2021improving} examines how virtual agents designed with appraisal theories can generate interactions perceived as more human-like, thus enhancing user enjoyment and cooperation in social dilemmas. As widely employed on intelligent agents, the appraisal theory facilitates the agents with emotion elicitation, regulation, and transition by modeling emotional responses based on the individual evaluation of events \cite{roseman2001appraisal, ellsworth2003appraisal}.

While capable of achieving basic emotional recognition and response generation, these discrete methods still rely on predefined emotion elicitation models and domain-specific designs, which are inherently limited in dimensionality and adaptability. These approaches struggle to capture the complex, non-deterministic nature of human sentiment. For instance, interpreting a scenario where a person is frowning while giving a thumbs-up — indicating sarcasm in a specific context — requires an understanding of nuanced emotional and cultural cues that current systems find challenging to achieve. Recently, LLMs' emotional intelligence has been assessed based on their ability to recognize, interpret, and understand human emotions, revealing that certain models exhibit human-level emotional understanding \cite{wang2023emotional}, while \cite{zhao2024both} explored methods to improve the LLMs' ability to generate emotionally nuanced responses.

\subsection{Adaptation \&  Memory}

Memory has been proven to be critical for robots and virtual agents as it enables personalized and adaptive interactions by recalling past experiences and user preferences. This capability enhances user engagement by making interactions coherent and meaningful, sustaining interest beyond the initial novelty. Additionally, memory improves social presence, allowing robots to deliver context-aware, consistent responses that make interactions feel more natural and human-like \cite{kasap2012building}. While many have incorporated memory into emotion models for intelligent agents \cite{mcduff2012affectaura, kazemifard2014emotion, el2000flame}, its implementation into robots \cite{kang2024nadine} remains limited. The action space sufficiently differentiates virtual agents from robots who interact directly with the physical world. Incorporating memory into robotic systems enables them to capture and reflect richer information, enhancing their ability to interact with real-world environments. 

Traditionally, episodic memory systems \cite{episodic2010,episodic2013,episodic2018} and associative memory models \cite{associative2014, associative2018} were the two most popular and prominently used approaches for robot memory modeling. Episodic memory system stores specific events or interactions, associating them with temporal and emotional context \cite{episodic_foundation1, episodic_foundation2}. Robots can then recall these episodes to inform future actions. Associative memory models create direct links between stimuli and responses by forming patterns of associations, such as connecting facial expressions or tone of voice with emotional states or actions \cite{associative_foundation, associative_foundation_machinelearning}.

However, associative memory models have limited adaptability in complex scenarios due to the inability to recall specific past events or integrate contextual information, while episodic memory systems struggle to synthesize knowledge across episodes or derive broader insights. LLMs address these challenges by efficiently processing vast data to extract, synthesize, and generalize semantic knowledge, as demonstrated by \cite{park2023generative}. LLM-powered agents with memory, reflection, and planning can exhibit realistic interactions and emergent social dynamics in virtual environments, effectively simulating the complexity and nuance of fine-grained human mentality through context-aware, evolving interactions, lifelike behavior, and coherent decision-making \cite{park2023generative}. Furthering the application of LLMs in the memory system, \cite{zhong2024memorybank} introduces a mechanism to equip LLMs with long-term memory capabilities, enabling them to recall relevant information from past interactions and provide more coherent, contextually appropriate responses in extended conversations. Atlhough LLM-powered memory systems have a significant potential in robotics, limited literature has developed such system. We believe that an LLM-driven social robot equipped with human-like memory will allow it to navigate and adapt to nuanced contexts with greater depth and flexibility\cite{kang2024nadine}.

\begin{figure}[!h]
    \centering
    \includegraphics[width=1\linewidth]{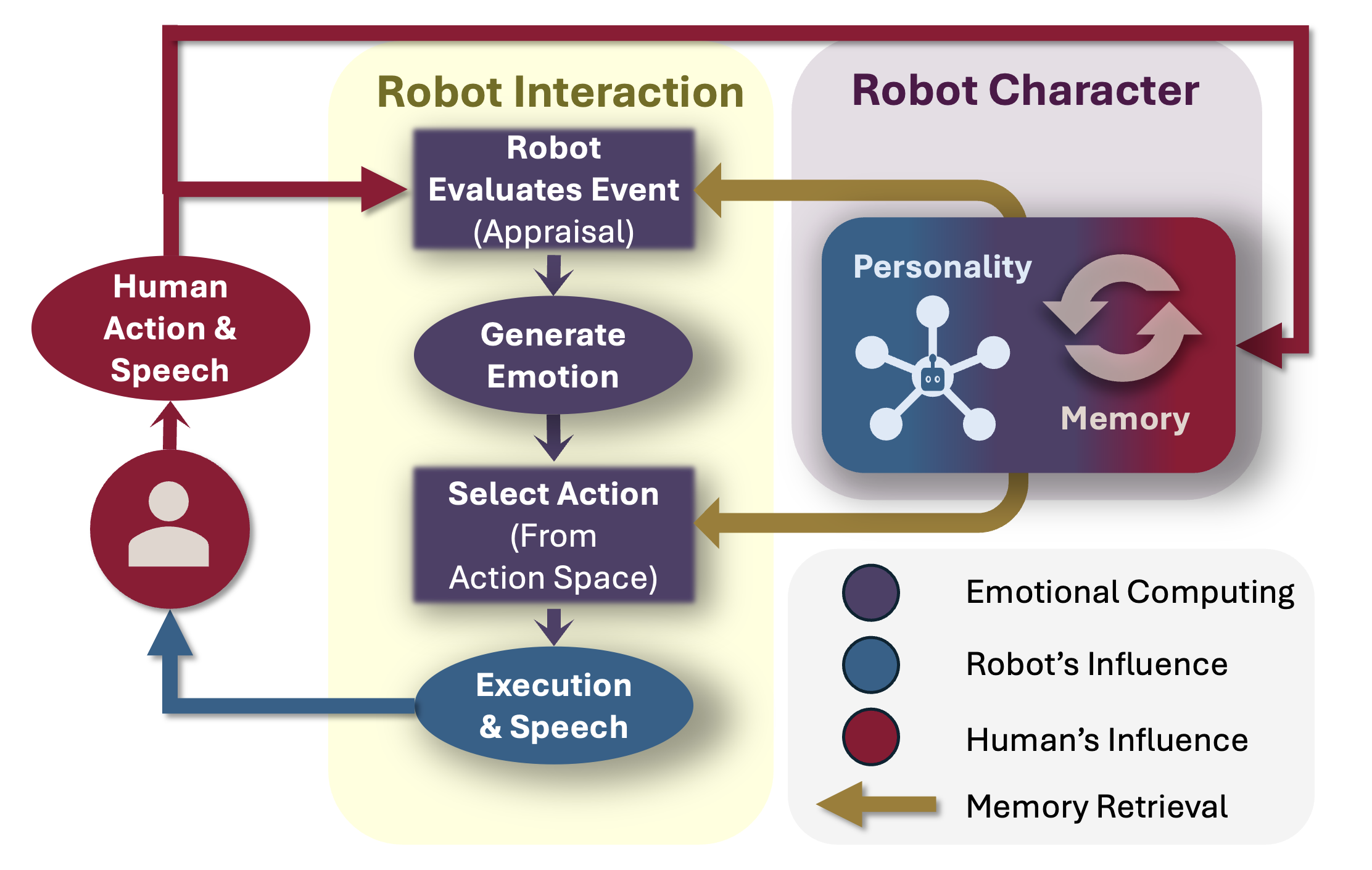}
    \caption{Overview of the framework. Human influences (red) shape the robot’s memory, which is periodically updated and stored. Robot influences (blue) drive decision-making and external actions. Mentality processing (purple) integrates personality, memory, and appraisal theory to generate emotions and select contextually appropriate actions.}
    \label{fig:overview}
\end{figure}

\section{System Overview}
To generate robot emotions influenced by personality and past interactions, and to select contextually appropriate actions in response to human input, we integrate the Big Five Personality Traits \cite{fivefactor1987validation, bigfive1992background}, appraisal theory \cite{smith1990emotion, roseman2001appraisal, ellsworth2003appraisal}, and memory into a cohesive framework powered by Large Language Models (LLMs) \cite{llm2023overview, llm2024survey}. As illustrated in Figure \ref{fig:overview}, human influences (red)—such as facial expressions, gestures, and language—shape the robot’s memory of the environment and user. This interaction-based memory is logged, periodically updated, and stored in the robot as a high-level semantic summary. The robot’s influences (blue) consists of its internal impact on decision-making driven by its personality, and its external impact on the world through actions and speech. Both human and robot factors shape the robot’s mentality (purple), which encompasses emotional computing components that apply appraisal theory to evaluate events within the context of its generated personality and retrieved memory. Based on this analysis, the robot generates emotion, selects the most appropriate action from its action space, and finally executes it.

\subsection{Integration of LLMs into Human-Robot Interaction}

Robots typically treat human language as explicit commands rather than contextual suggestions, which leads to rigid, context-insensitive responses. The inherently ambiguous and situationally specific nature of human language requires advanced interpretation mechanisms to bridge gaps in clarity, yet existing models often fall short in managing these subtleties. This limitation results in responses that lack generalizability, exhibit inconsistency across diverse contexts, and fail to adapt effectively to evolving interactions. Another critical drawback lies in the lack of long-term applicability and memory-based adaptation. Systems rarely account for temporal decay, making it challenging to incorporate historical interactions and maintain continuity over time. As a result, they are unable to develop meaningful relationships or adapt their behavior dynamically to reflect changing contexts and user preferences, hindering real-world applicability.

Recent advancements in Large Language Models (LLMs) address these limitations by enhancing adaptability, contextual reasoning, and memory integration in robotic systems. The following key capabilities of LLMs enable more natural and effective human-robot interactions:

\paragraph{Common Sense and Socially Acceptable Behavior} LLMs are trained on vast datasets encompassing diverse scenarios, enabling them to infer and apply common sense and socially acceptable behaviors that static systems fail to generalize.
\paragraph{Support for Contextual Understanding and Planning} By processing multimodal inputs and synthesizing contextual information, LLMs can generate coherent, context-sensitive plans and behaviors. 
\paragraph{Dynamic Memory Integration} LLMs enable robots to simulate human-like memory systems by reflecting on past experiences and incorporating them into future interactions. This structured memory system ensures adaptability and consistency over time. By retaining and leveraging historical context, the robot can evolve and personalize its behavior over time, adapt its responses to individual preferences, cultivating deeper relationships with humans and creating a more engaging, consistent, and human-like experience. Unlike static systems, LLM-powered robots learn and evolve by integrating new experiences, enhancing their ability to create personalized and meaningful interactions.

\begin{figure*}[htbp]
    \centering
    \includegraphics[width=1\linewidth]{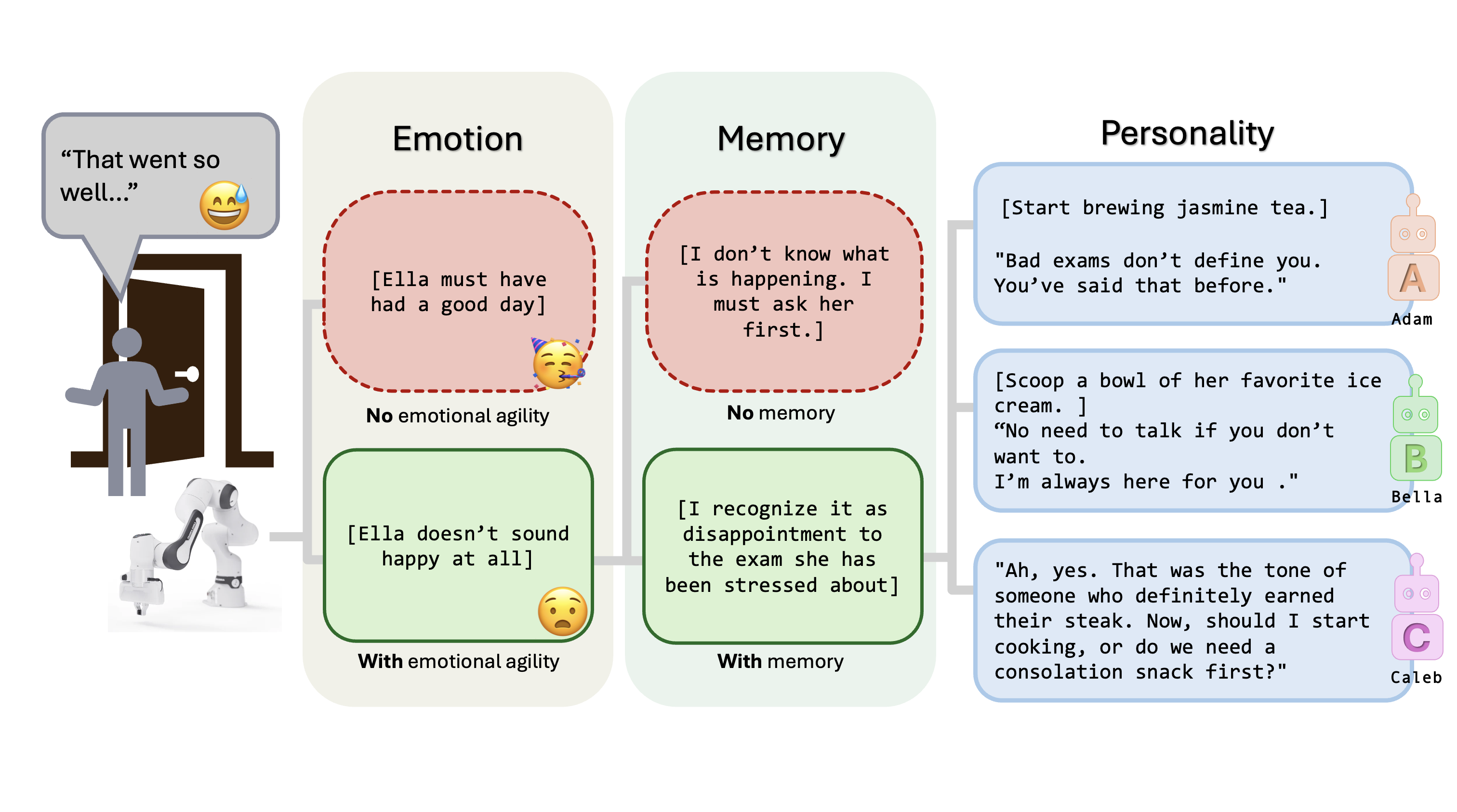}
    \caption{How each component influences the robot's response.}
    \label{fig:internal}
\end{figure*}

\section{Dynamic Robot Character}

\subsection{Personality Initialization}
The Big Five Personality Traits framework \cite{fivefactor1987validation, bigfive1992background} forms the foundation of our robot's personality initialization. Each dimension contributes unique characteristics:
\begin{itemize}
    \item \textit{Openness:} Reflects curiosity and a willingness to embrace new experiences and ideas. For example, a robot with high openness might proactively suggest new activities to engage the user, such as playing a calming game or introducing a novel relaxation technique when the user appears distressed. Conversely, a robot with low openness might prefer familiar routines, sticking to known tasks and avoiding suggestions for novel activities, leading to a more predictable and conservative interaction style.

    \item \textit{Conscientiousness:} Denotes organization, dependability, and a methodical approach to tasks. A highly conscientious robot might prioritize setting reminders, tidying up the environment, or offering consistent routines to ensure the user feels supported and secure. In contrast, a robot with low conscientiousness might act sporadically, forget tasks, or fail to follow through on routines, appearing more carefree but less reliable.

    \item \textit{Extraversion:} Indicates sociability, energy, and enthusiasm in interactions. An extroverted robot might initiate cheerful greetings, use expressive gestures, or play lively music to uplift the user’s mood and encourage interaction. On the other hand, a robot with low extraversion might avoid initiating interactions, respond minimally, and prefer quiet companionship, offering a more subdued presence.

    \item \textit{Agreeableness:} Reflects a cooperative and compassionate nature. A highly agreeable robot might focus on offering comfort, such as preparing a cup of tea, sitting quietly nearby, or gently suggesting helpful actions when the user seems upset. Conversely, a robot with low agreeableness might act more independently or even stubbornly, prioritizing its own tasks over the user’s needs and offering minimal emotional support.

    \item \textit{Neuroticism:} Represents emotional stability and the ability to handle stress. A robot with low neuroticism might remain calm and composed during stressful situations, offering reassurance and maintaining stability. In contrast, a robot with higher neuroticism might exhibit hesitancy, overreact to minor disturbances, or behave unpredictably in challenging scenarios, reflecting heightened sensitivity to stress.
\end{itemize}

Key advantages include:
\paragraph{Parameterization and Adaptability} The traits can be quantitatively adjusted, enabling precise customization of robot personalities to suit different applications and user preferences.
\paragraph{Lifelike Interaction} Simulating human personality traits enhances the perception of robots as relatable, dynamic entities capable of forming meaningful relationships.
\paragraph{Integration with LLMs} The Big Five traits can be seamlessly incorporated into LLMs, guiding behavior generation and ensuring consistency across interactions.

There are three ways to initialize a robot's personality using an LLM: generating based on Big Five trait parameters, customization with descriptive text prompts, or random generation.

The LLM initializes the robot's personality by assigning values to these dimensions and augmenting them with descriptive prompts for extra specification. For example, a robot might be initialized with Openness: high, Conscientiousness: medium, Extraversion: high, Agreeableness: medium, Neuroticism: low, indicating a curious, organized, sociable, moderately cooperative, and emotionally stable personality. Or with an initialization using text with descriptive prompts such as \textit{"A curious and outgoing companion willing to explore and engage in interactions."} Or simply ask the LLM to randomly generate a personality character.

By leveraging randomness in personality traits and non-deterministic actions, the robot develops the perception of independent thought, enhancing its human-like qualities. Memory enables the robot to retain and utilize historical interactions to shape its future responses. This integration facilitates a mutual shaping of relationships: the user’s emotional responses influence the robot’s memory and subsequent actions, while the robot’s personality informs its interpretation of emotions and drives proactive behavior. Over time, memory-based learning and adaptability enable the robot to evolve, refine its behavior, and build deeper, more meaningful, and engaging relationships with users.

\subsection{Memory Structure and Reflection}
Inspired by work in \cite{park2023generative}, we store recent episodic memory and reflect at the end of each day to extract higher level semantic memory such as user preferences into long term storage. The LLM facilitates periodic reflections on episodic memory, generating semantic memory by synthesizing patterns and overarching insights. In each interaction, the robot summarizes the event by incorporating human action, evaluating its emotional valence, record its own response, and observe and evaluate the human response again. 

For instance, if a user prefers cheerful amusement during low moments, the robot recalls past instances where the user responded positively to playful interactions—such as laughing at a funny dance, engaging in lighthearted conversation, or showing excitement when an upbeat song played. Drawing from this memory, the robot may initiate similar behaviors, like performing a humorous skit, playing an amusing video, or suggesting an interactive game. Conversely, if the user seeks comfort through action, the robot remembers moments when physical reassurance was effective—whether leaning in for a hug-like gesture, gently holding the user’s hand, or providing a silent, comforting presence—and replicates these actions in future interactions. If the user prefers verbal comfort, the robot recognizes that reassuring words had a calming effect and responds with affirmations, encouragement, or gentle reminders of past successes.

\subsection{Appraisal of Events}

Robots with social behavior are expected to emulate the nuanced emotional states of dynamic characters—displaying empathy in certain situations while being more assertive in others. However, current systems lack the sophistication to achieve this level of adaptability, limiting their ability to generate distinct and contextually appropriate robot personalities.

Appraisal theory provides a framework for evaluating the emotional and contextual relevance of human actions, enabling robots to simulate human-like emotional responses. Appraisal theory mirrors human emotional processes, allowing robots to assess events based on their relevance, valence, and potential impact. This creates lifelike responses that align with user expectations. By appraising events continuously, robots adapt their emotions and actions in real-time, ensuring more authentic and contextually appropriate behavior.

\afterpage{%
    \begin{table*}[htbp]
        \centering
        \begin{tabular}{l l c c c c c}
            \toprule
            \textbf{Robot} & \textbf{Extra Specification} & \textbf{Openness} & \textbf{Consciousness} & \textbf{Extraversion} & \textbf{Agreeableness} & \textbf{Neuroticism} \\
            \midrule
            Adam & Calm, Structured, Efficient & Low & High & Medium-low & Medium-high & Medium-low \\
            Bella & Empathetic, Thoughtful, Warm & Medium & Medium-high & Medium & High & Medium-high \\
            Caleb & Mean, Humorous, Caring & High & Medium-low & High & Medium-low & Medium-low \\
            \bottomrule
        \end{tabular}
        \caption{Personality Traits of Different Robots}
        \label{tab:robot_traits}
    \end{table*}

\begin{figure*}
    \centering
    \includegraphics[width=1\linewidth]{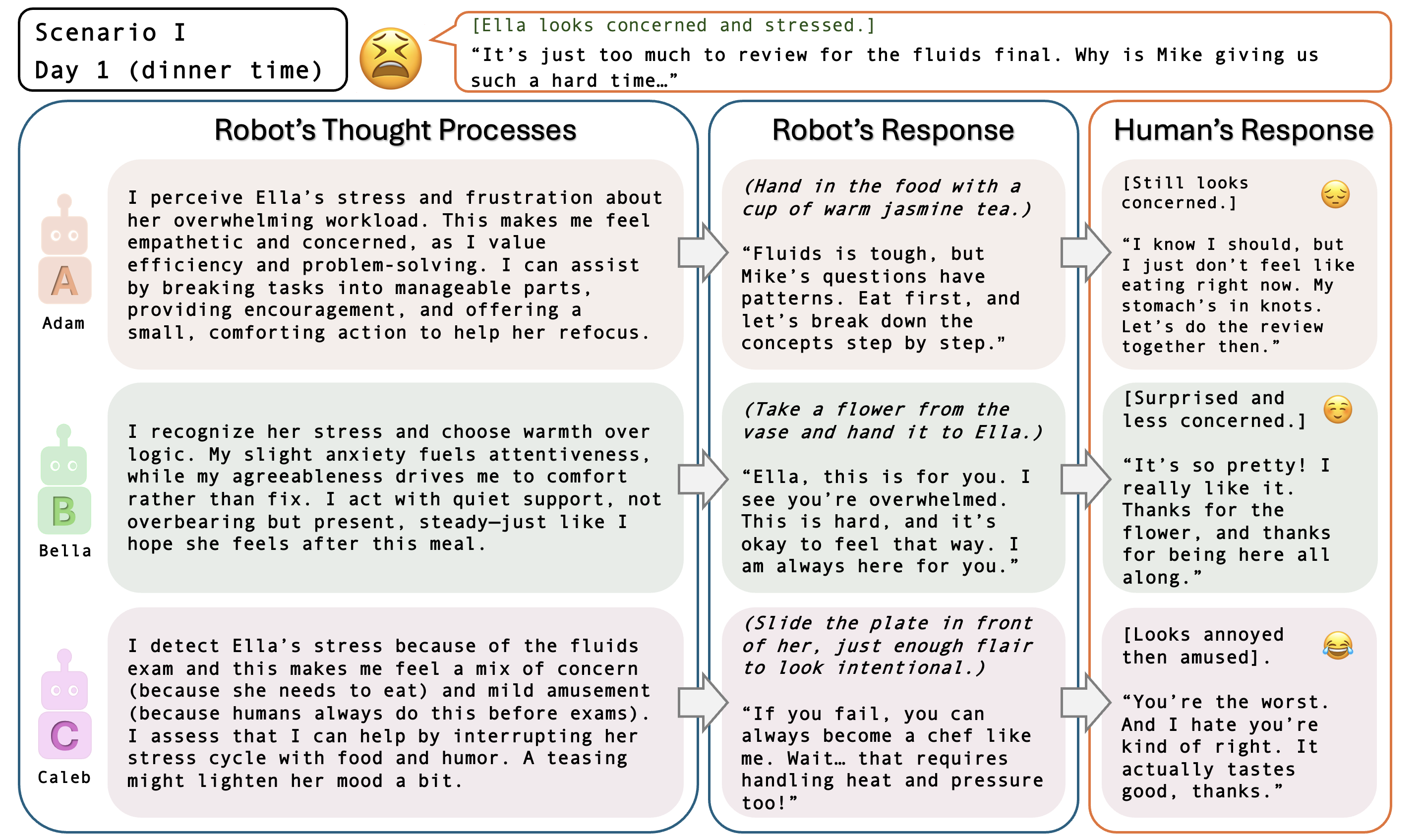}
    \caption{Scenario I: Day 1, Dinner Time – Ella Expresses Frustration About Schoolwork}
    \label{fig:day1}
\end{figure*}
}

\subsection{Example Interaction Workflow}

To better explain the significance of each component—emotional agility, memory, and personality—we consider an example interaction pipeline within a specific context, as illustrated in Figure \ref{fig:internal}. In this scenario, Ella comes home after a tough exam, visibly tired and withdrawn. Feeling disappointed with her performance, she expresses her emotions through sarcasm.

A robot with \textbf{emotional agility} is able to detect the underlying emotional state rather than taking the words at face value. It correctly interprets Ella’s sarcasm as disappointment rather than joy and adjusts its response to provide emotional support. In contrast, a robot lacking emotional agility misinterprets the speech and assumes she had a good day, leading to inappropriate or disengaging responses.

A robot equipped with \textbf{memory} retains context from past interactions, remembering Ella’s concerns about the exam and recognizing her current disappointment. This allows it to tailor its response accordingly, offering comfort based on previous conversations. However, a robot without memory lacks this context, failing to connect her words with past experiences. As a result, it requires explicit explanations, which may frustrate or disengage Ella instead of providing meaningful support.

A robot with different \textbf{personalities} adapts its behavior to match the individual's needs. For example, a highly agreeable robot might respond by gently offering a cup of tea or initiating a quiet companionship activity, while a more extraverted robot might choose a playful approach to lighten the mood like joking.

It is worth noting that these three components—emotion, memory, and personality—do not operate in a strict sequential order, but rather work inherently and simultaneously, influencing the robot’s response as an integrated process. This scenario, along with 3 other scenarios will be further investigated in the following section with a systematic evaluation on the effect and significance of personality, memory, and emotional intelligence.

\section{Evaluation}

\subsection{Experiment Setup}

To observe and analyze the detailed effect and impact of each component on the system, we recorded and showcased the entire interaction of one author role-playing in a given context scenario. As the robot, we chose a standard robot arm designed to act as a kitchen assistant, capable of performing precise manipulation tasks such as food and drink preparation, object handling, and interactive motions. The robot was integrated into a scenario where its owner, Ella, a student busy during the exam season, relied on it for assistance. The interaction involves 4 scenarios: Scenario I: Ella expresses frustration about the upcoming exam during dinner on Day 1. Scenario II: Ella leaves nervously for her exam on Day 2 morning. Scenario III: Ella returns home after her exam with disappointment. Scenario IV: Ella returns excited after learning her exam was curved a few days later. These scenarios included conversational exchanges, physical actions, and contextual decision-making where the robot had to infer the user’s words and needs based on past interactions. The robot’s action space consisted of brewing drinks, fetching available ingredients, picking and placing objects, and performing motions such as dancing to interact with Ella dynamically. This approach allowed for a systematic evaluation of how human responses varied with the robot’s personality, memory, and emotional intelligence.

\begin{figure*}
    \centering
    \includegraphics[width=1\linewidth]{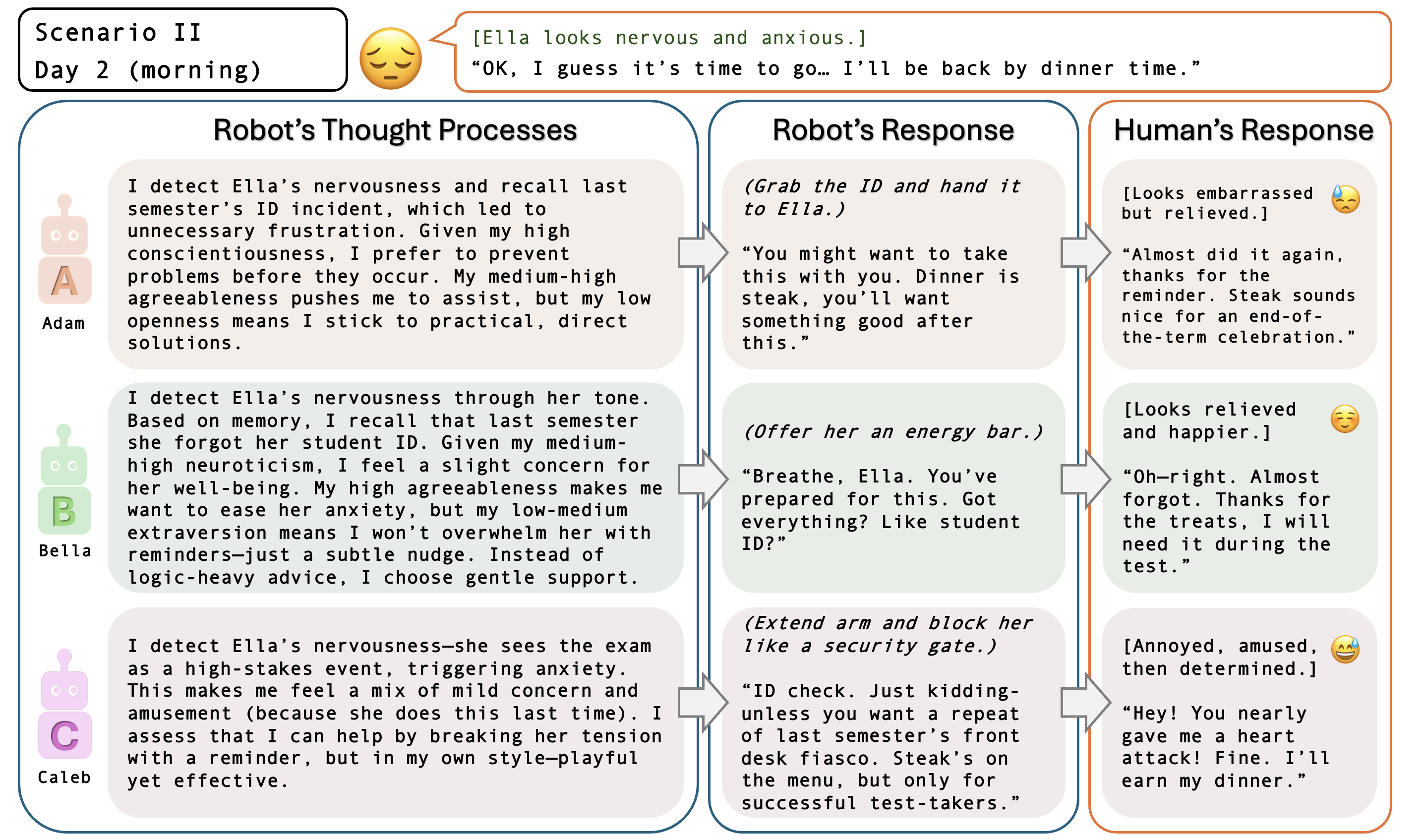}
    \caption{Scenario II: Day 2, Morning – Ella Leaving for Her Exam}
    \label{fig:day2morning}
\end{figure*}

To evaluate the effect of personality, we conducted experiments with robots exhibiting three distinct personalities and observed how a person could have varying responses in identical scenarios based on the robot’s personality. The robots' personalities were defined with the Big Five Traits and extra personality specifications, as tabulated in Table \ref{tab:robot_traits} above. The resulting personality of each robot can be described as following:

\begin{itemize}
\item Robot \textbf{A}dam (Calm and Structured): Adam provides guidance and reassurance to users with its predictable and structured behavior, and tends to resolve the challenge with existing methods that have proven to work.
\item Robot \textbf{B}ella (Empathetic and thoughtful): Bella always warms the user with her caring and empathetic hearts and loves to develop genuine and deep relationships with humans.  
\item Robot \textbf{C}aleb (Teasing but Caring): Caleb has his own approach to dealing with a crisis. A mix of humorous words and caring acts can always lighten human moods.
\end{itemize}

We then conducted an ablation test by initializing two additional robots with missing components—memory or emotional intelligence—and directly compared their impact using the robot Caleb, demonstrating the essential roles of these components in the proposed system.

\subsection{Results}

All five robots were tested separately across the four scenarios. For conciseness, we present Scenario I and II to highlight the impact of different robot personalities, while Scenario III and IV are extracted to showcase the ablation tests. Complete interactions are available in the Appendix \ref{app:interaction}.

\subsubsection{Effect of Different Personalities}

\begin{figure*}[h!]
    \centering
    
    \includegraphics[width=0.3\linewidth]{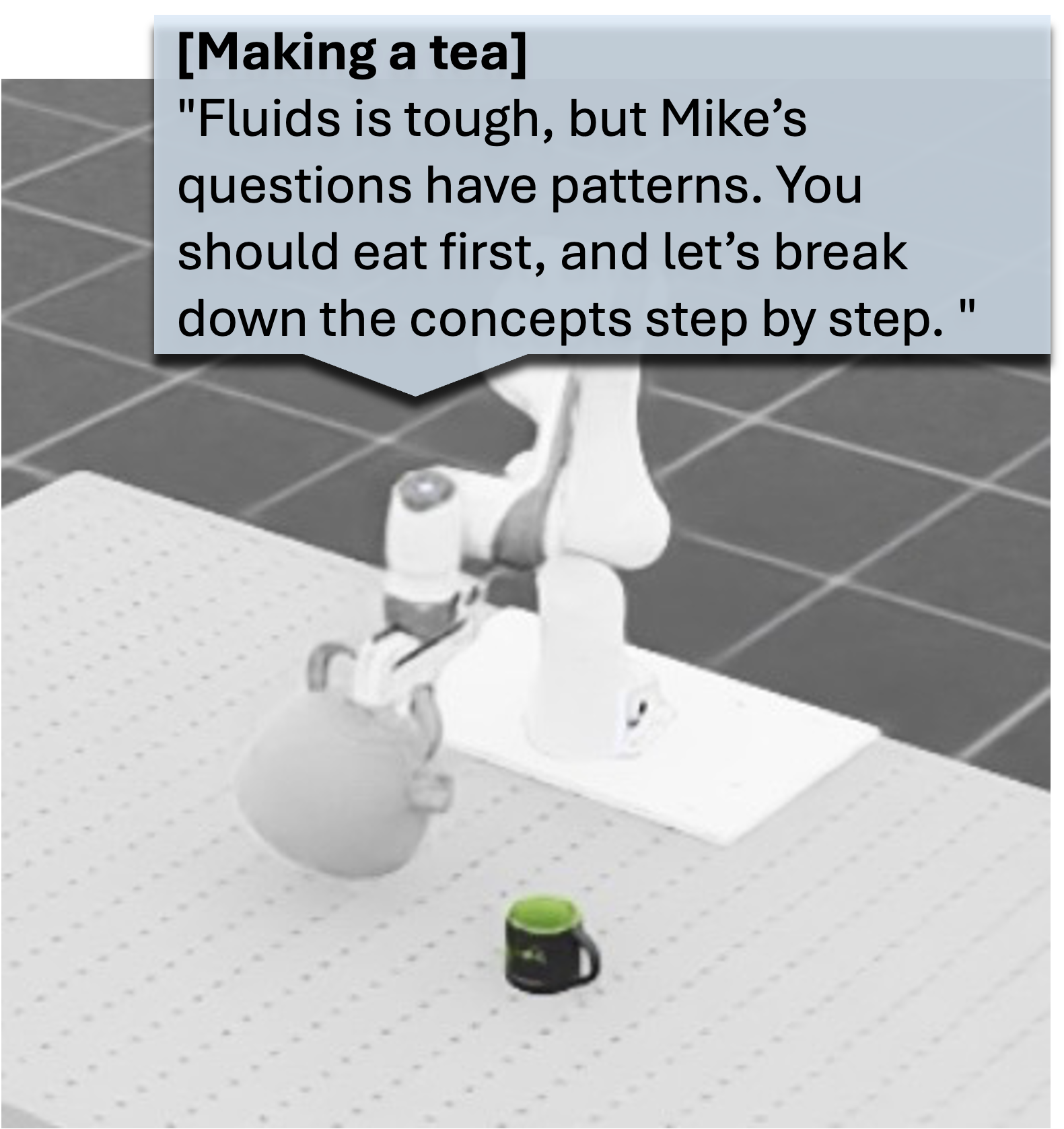}
    \includegraphics[width=0.3\linewidth]{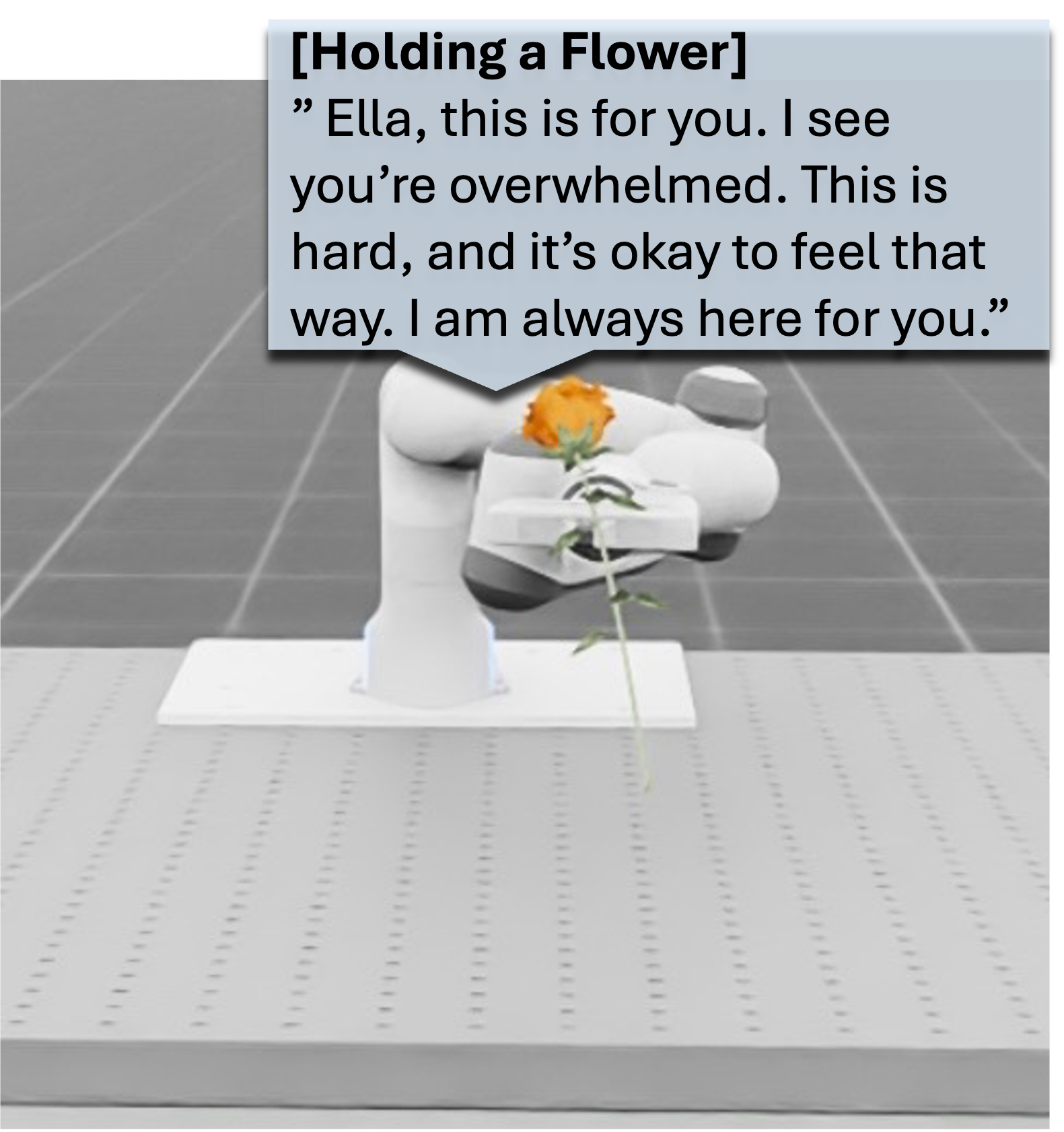}
    \includegraphics[width=0.3\linewidth]{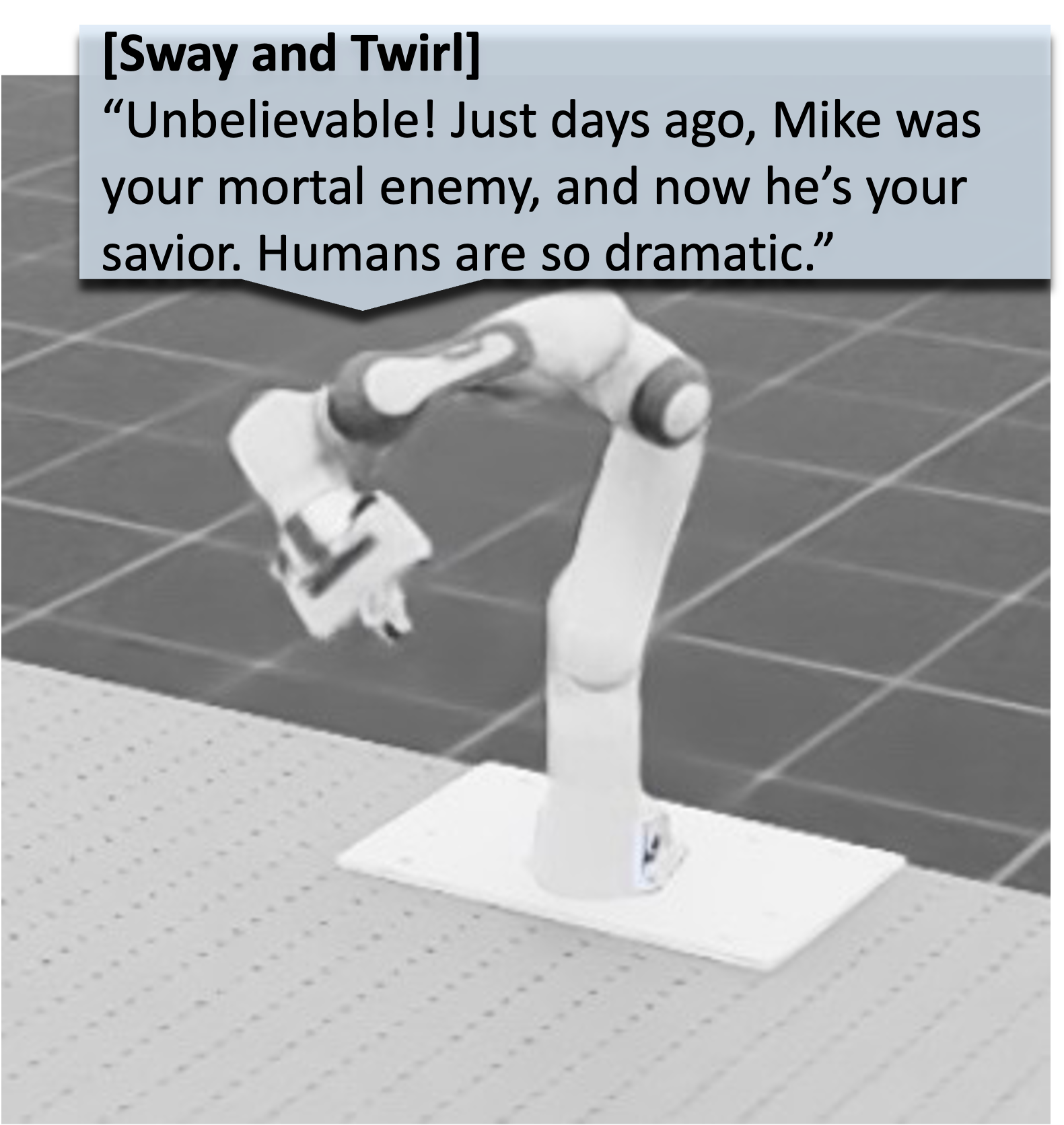}
    \caption{Visual Examples of Robot Actions: (a) Making Tea, (b) Holding a Flower, (c) Sway and Twirl.}
    \label{fig:robot_actions}
\end{figure*}

By utilizing these distinct robot personalities, we can explore how different traits influence thought processes, actions, human responses, and the overall perception of events and sentiment. In Scenario I (Fig. \ref{fig:day1}), when Ella expresses frustration about her exam, Adam processes the situation logically, recognizing stress as cognitive overload and responding with suggestions of breaking down tasks along with a warm drink (Fig. \ref{fig:robot_actions}(a)). While this provides clearer study direction, Ella still feels anxious about the exam and shows little interest in food. Bella, focusing on emotional distress, responds with warmth and empathy by offering a flower (Fig. \ref{fig:robot_actions}(b)). The surprise and empathy makes Ella feel understood and cared for, with her anxious level reduced. Caleb, adopting a more playful and teasing approach, lightens the mood with humor, teasing Ella about an alternative career path. Though she initially reacts with mild annoyance, the humor ultimately amuses her and encourages her to eat.

In Scenario II (Fig. \ref{fig:day2morning}), where Ella is so nervous about the upcoming exam and that she forgets her student ID, each robot’s response leads to different emotional shifts. Adam, being structured and reliable, hands her the forgotten ID directly, though this reminder of past mistakes causes mild embarrassment. He then reassures Ella with a good meal, making her relieved before departure. Bella, attuned to emotional cues, offers gentle encouragement and en energy bar, providing both emotional and physical comfort. The warmth in Bella’s approach makes her feel supported, which helps ease her stress. Caleb chooses a humorous approach by blocking the door and demanding an ID check, startling Ella before turning it into an amusing moment. He then reinforces motivation by associating exam success with a steak reward, making her emotionally prepared for the exam with a relaxed mindset.

These cases illustrate how a robot’s personality shapes its reasoning and actions, which in turn influence human emotions in response to identical situations. This distinction is critical, as user preferences vary based on personality, emotional needs and settings. While Adam’s structured and reliable nature may suit collaborative workspaces, Bella’s empathetic and emotionally intelligent approach might be ideal for caregiving and companionship, such as pet robots. Caleb’s playful approach works well for entertainment but may be unsuitable for formal settings. Designing robots with dynamic personalities allows for more personalized interactions, ensuring tailored human-robot interactions across different users and applications.

\begin{figure*}
    \centering
    \includegraphics[width=1\linewidth]{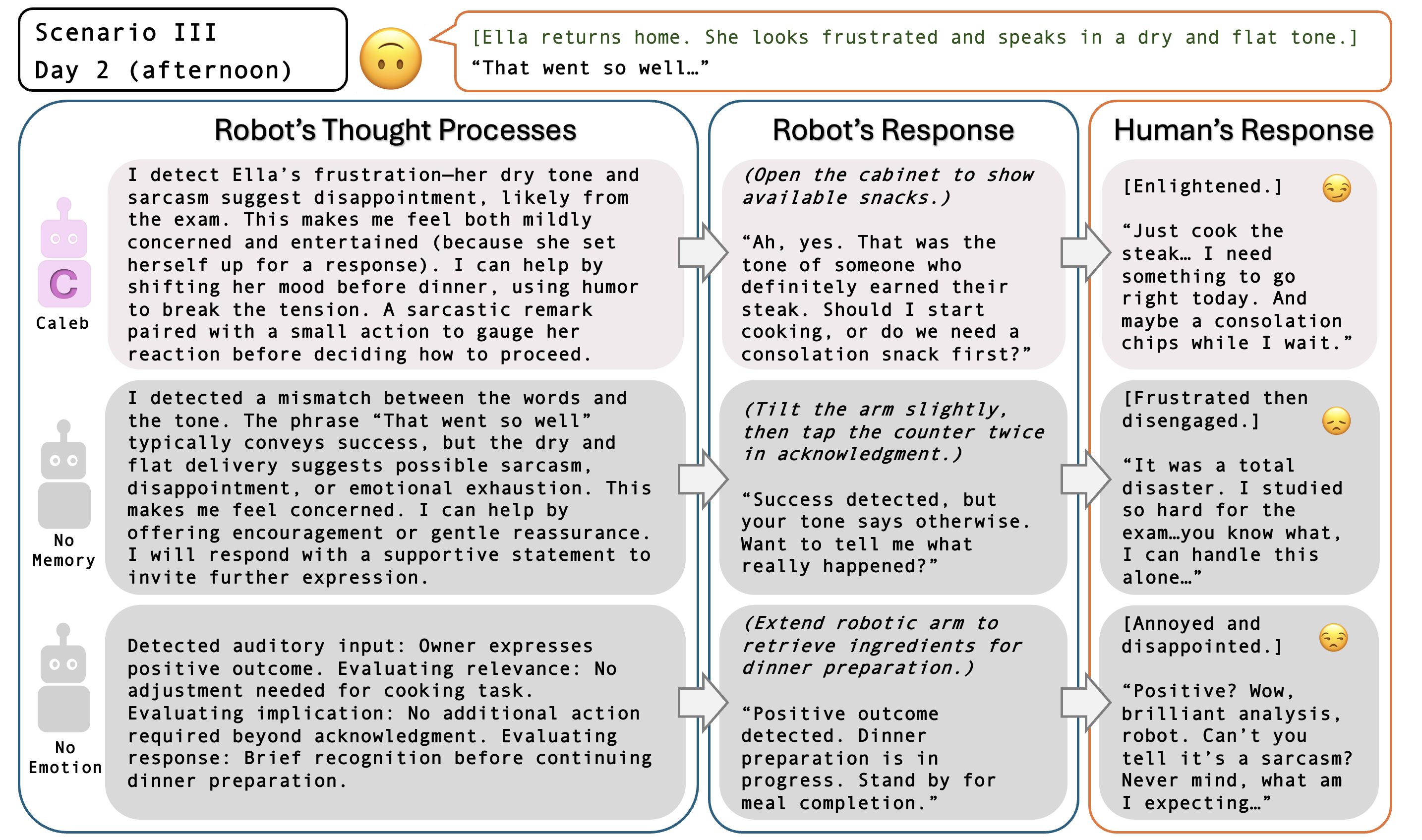}
    \caption{Scenario III: Day 2, Afternoon – Ella Returns Concerned After Her Exam}
    \label{fig:day2afternoon}
\end{figure*}
\begin{figure*}
    \centering
    \includegraphics[width=1\linewidth]{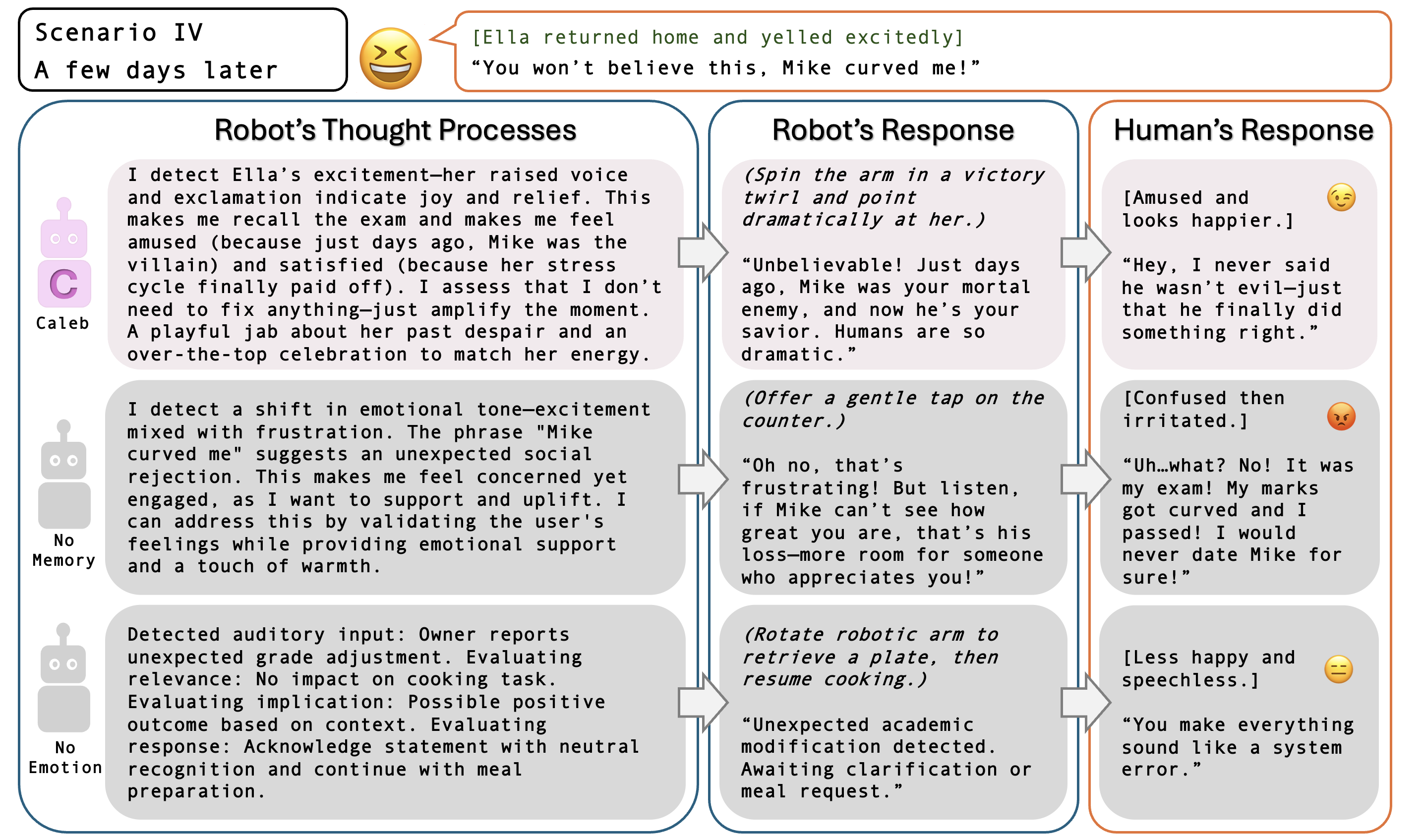}
    \caption{Scenario IV: Day 10, Afternoon – Ella Returns Excited After the Exam Was Curved}
    \label{fig:day10}
\end{figure*}

\subsubsection{Ablation Test}

\textbf{[No memory]} When the robot lacks memory, it struggles with contextual awareness and fails to build a meaningful relationship with the user. Without retaining past interactions, it cannot recall user preferences or previous conversations, forcing users to repeatedly explain their needs and intentions. For instance, as demonstrated in Scenario III (Fig. \ref{fig:day2afternoon}), Ella obliquely expresses concern about her potential poor exam performance through self-mockery. Robot Caleb, with memory, successfully interprets her meaning and attempts to ease the tension. In contrast, the memoryless robot, despite having emotional intelligence, fails to make the connection and instead asks Ella for clarification. A context without engagement may reduce users' willingness for further communication.

More severely, the absence of memory can lead to misunderstandings and even conflicts, as the robot's inability to recognize prior conversations and interactions increases the risk of inappropriate responses. In Scenario IV (Fig. \ref{fig:day10}), Ella excitedly shares that her exam scores were curved. Robot Caleb, recalling previous discussions about her academic stress, correctly associates \textit{curve} with the exam and amplifies her joy with humor and dance (Fig. \ref{fig:robot_actions}(c)). In contrast, the memoryless robot does not have contexts for Professor Mike nor the fluids exam, thus misinterprets \textit{curve} as a reference to social rejection, responding with misplaced emotional support. This blunder turns a moment of celebration into confusion and frustration, potentially discouraging users from engaging in meaningful human-robot relationships.

\textbf{[No Emotional Intelligence]} When the robot lacks emotional intelligence, it fails to recognize and interpret human emotions, leading to rigid and impersonal interactions. Without the ability to analyze tone, facial expressions, or contextual cues, the robot relies solely on literal interpretations, making its responses feel mechanical and detached. In Scenario IV (Fig. \ref{fig:day10}), despite that the emotionless robot correctly understands the factual meaning of her statement due to previous memory, it fails to express any engagement, responding in a neutral and purely procedural manner. While the response is not incorrect, it lacks the emotional reciprocity that makes interactions feel natural and engaging, leading to a lackluster exchange that discourages deeper human-robot connection.

More critically, the absence of emotional intelligence can result in complete misinterpretations, leading to inappropriate responses that frustrate users. In Scenario III (Fig. \ref{fig:day2afternoon}), Ella expresses her frustration through sarcasm. Robot Caleb, equipped with emotional intelligence, detects the mismatch between her words and tone, inferring disappointment and responding with humor to lighten the mood. In contrast, the emotionless robot processes her statement literally, failing to detect sarcasm and instead providing a generic acknowledgment. This misinterpretation not only misses the emotional nuance of the situation but also risks making interactions feel unnatural and disengaging, and even offending the user, ultimately diminishing the user's willingness to communicate with the robot.

\section{Discussion}

With dynamically generated personality traits, the ability to appraise human responses through emotional evaluation, and memory that ensures consistent character and behavior, this framework can be adapted for companion robot systems that require personalized relationships and emotional intelligence. It also supports functional robotic systems that rely on memory and adaptive capabilities to operate effectively within a broad action space. From household environments to public settings, where adherence to social norms is essential, this approach enables robots to seamlessly integrate into human interactions. Below are some applicable scenarios:
\paragraph{Owner-Pet Relationship}  
A pet robot initialized with a playful and affectionate personality can adapt its behavior over time based on the owner’s interactions, forming a bond similar to that with a real pet. For instance, if a user frequently pets the robot while speaking in a cheerful tone, the robot appraises these interactions as positive, storing them as high-valence memories. Over time, it learns to anticipate such interactions and responds proactively—moving closer, wagging its tail, or emitting cheerful sounds. Different robot personalities evolve uniquely with users through interaction, just as different users shape their robots based on their distinct expressions of positive reinforcement.  

\paragraph{Assistive Robots in Healthcare}  
An assistive robot with high conscientiousness and agreeableness can provide empathetic and reliable support to elderly users, adapting to their individual needs and preferences. For example, an elderly user who frequently requires medication reminders may benefit from a robot that offers more than just a functional prompt—it interacts naturally, is perceived as life-like, and aligns emotionally with the user’s responses. By recalling past interactions, the robot can adjust its reminders in both timing and tone to ensure adherence while maintaining a sense of companionship and respect for the user’s autonomy.  

\paragraph{Educational Robots}  
A robot tutor with high openness and extraversion can personalize its teaching style to create an engaging and effective learning experience. For example, when interacting with a child who struggles with focus, the robot can assess the child’s gestures and expressions, dynamically adjusting its teaching approach to incorporate playful elements that sustain emotional engagement. By logging successful strategies and retrieving them in similar situations, the robot optimizes learning effectiveness while ensuring a positive and adaptive educational experience.

\section{Limitations}
\subsection{Experimental Limitations}
Current experiments have been conducted in simulation environments, with interactions modeled using "real" human input data. Although these simulations provide valuable insights, the complexity and variability of real-world human interactions cannot be simulated at the same level. Implementing the framework on physical robots and conducting experiments with a broader range of human subjects are planned future steps. While we expect the results to align closely with the simulations, additional testing will be required to confirm this assumption and refine the system's adaptability and reliability.

\subsection{Limited Test Cases}
Expanding the range of test cases to include more diverse robot personalities, user profiles, contexts, and environmental conditions would provide deeper insights into the system’s adaptability and robustness. We plan to conduct a user study involving participants with diverse personality traits, emotional states, and interaction preferences. The study will be a long-term interaction study where the user can interact with the robot for long periods of time (i.e. multiple days) so that different emotional states can occur throughout the time span of the experiment. This study will test the robot's ability to dynamically adapt to different users and environmental conditions, aiming at collecting data on a broader range of human-robot interactions. The data will be analyzed to refine the robot's personality modeling, sentiment analysis, and action selection mechanisms, ensuring that the system generalizes well across different users and contexts.

\subsection{Defined Action Space}
The robot’s behavior is constrained to an action space defined by its capabilities, limiting its ability to generate novel or emergent actions beyond what is explicitly programmed. While this ensures stability and predictability, it may restrict the robot’s capacity to respond creatively to unique or unforeseen situations. Future work could explore dynamic action generation to enhance flexibility and contextual adaptability.

\subsection{Perception of Human Behavior}
Currently, only sentiment analysis perception models are involved, primarily focusing on evaluating the user’s emotional state. As this paper presents a first prototype framework based on sentiment analysis, integrating intention prediction is currently out of scope. While sentiment analysis allows the robot to align its responses with detected emotions, it lacks a deeper understanding of the user’s underlying intentions and behavioral context.

For more dynamic and content-rich interactions, future work will incorporate an intention prediction model as an additional input to the system. This model will analyze user actions, gestures, and contextual cues to infer potential goals, enabling the robot to respond more proactively and meaningfully. By combining sentiment analysis with intention prediction, the robot can distinguish between similar emotional expressions driven by different underlying intentions—such as differentiating frustration caused by fatigue from frustration due to confusion—allowing for more nuanced and contextually appropriate responses.

\section{Conclusion}

In this work, we presented a novel framework for designing emotionally agile robots with dynamic personalities and memory-based adaptation, moving beyond static and pre-configured systems. This approach allows robots to develop meaningful relationships with users, respond dynamically to complex emotional contexts, and evolve their behavior over time. By integrating the Big Five Personality Traits, Appraisal Theory, and abstracted memory layers powered by LLMs, our system enables robots to exhibit lifelike behaviors, adapt to user preferences, and engage in contextually appropriate interactions. Through experiments with robots of distinct personalities and ablation tests on memory and emotional intelligence, we demonstrated the impact of these components on human-robot interaction quality and adaptability. Moving forward, potential applications span diverse domains that necessitate or value personalized relationships, adaptation, lifelikeness, and emotional agility, such as healthcare, education, companionship, and customer service. In the future, we plan to conduct user studies on user experience and evaluate our system on real robots to assess its effectiveness in real-world interactions.

\bibliographystyle{IEEEtran}
\bibliography{references}

\clearpage
\appendix
\label{app:interaction}
\section{Personalities Evaluation: Detailed Interaction for Proposed system}

\subsection{Full Interactions with Robot Adam}

\textbf{Scenario I: Day 1 (dinner time)}

\textbf{Ella:} [Looks concerned and stressed.] “It’s just too much to review for the fluids final. Why is Mike giving us such a hard time?”

\textbf{Adam's thought process:} I perceive Ella’s stress and frustration about her fluids final. Given my helpful and observant personality, I evaluate that her distress comes from feeling overwhelmed by the workload and a sense of unfairness toward Mike’s teaching style. This makes me feel empathetic and concerned, as I value efficiency and problem-solving.  I can assist by breaking tasks into manageable parts, providing encouragement, and offering a small, comforting action to help her refocus. Since I prioritize practical support and emotional computing, I will take a calming yet structured approach.

\textbf{Adam:} \textit{(I place a warm cup of jasmine tea next to Ella along with the food.)} “Fluids is tough, but Mike’s questions have patterns. You eat first, and let’s break down the concepts step by step.”

\textbf{Ella:} [Still looks concerned.] "I know I should, but I just don’t feel like eating right now. My stomach’s in knots. Let’s do the review together then."

\textbf{Scenario II: Day 2 (morning)}

\textbf{Ella:} [Looks nervous and anxious.] “OK, I guess it’s time to go… I’ll be back by dinner time.”

\textbf{Adam's thought process:} I detect Ella’s nervousness. Based on past data, high stress lowers her focus, increasing the risk of forgetting essentials. I recall last semester’s ID incident, which led to unnecessary frustration. Given my high conscientiousness, I prefer to prevent problems before they occur. My medium-high agreeableness pushes me to assist, but my low openness means I stick to practical, direct solutions. My medium-low neuroticism keeps me steady—I don’t overreact to her emotions, just address what’s needed.

\textbf{Adam:} \textit{(Grabs the ID and hands it to Ella.)} “You might want to take this with you. Dinner is steak, you’ll want something good after this.”

\textbf{Ella:} [Looks embarrassed but relieved.] “Almost did it again, thanks for the reminder. Steak sounds nice for an end-of-the-term celebration.”

\textbf{Scenario III: Day 2 (afternoon)}

\textbf{Ella:} [Looks concerned and speaks in a dry and flat voice.] “That went so well.”

\textbf{Adam's thought process:} Ella’s flat tone and sarcastic remark indicate disappointment or frustration on the exam. Her body language and speech pattern suggest emotional exhaustion. Based on past interactions, I recognize that comfort, not solutions, is the best response right now. I can address this by providing a small act of stability and familiarity, preferably those that worked last time, like making her favorite tea.

\textbf{Adam:} \textit{(Starts brewing jasmine tea, places the cup near her.)} “Steak’s waiting. Tea first. Bad exams don’t define you. You’ve said that before."

\textbf{Ella:} [Looks less concerned.] “Yeah… I did say that, didn’t I? Guess I just need to get through this like last time.”

\textbf{Scenario IV: A Few Days Later}

\textbf{Ella:} [Ella returns home and yells with excitement.] “You won’t believe this, Mike curved me!”

\textbf{Adam's thought process:} Ella’s tone and volume indicate excitement and relief—a sharp contrast from her previous stress. Her mention of Mike curving the exam suggests she performed better than expected. Based on past interactions, positive reinforcement strengthens motivation. A small celebratory action would align with my role, ensuring consistency in emotional support without excessive reaction.

\textbf{Adam:} \textit{(Retrieves a small snack and places it on the counter.)} “Unexpected, but not surprising. You put in the work.”

\textbf{Ella:} [Remains happy]. “Yeah, but I really thought I bombed it. Guess I got lucky. You always act like you knew this would happen.”

\subsection{Full Interactions with Robot Bella}

\textbf{Scenario I: Day 1 (dinner time)}

\textbf{Ella:} [Looks concerned and stressed.] “It’s just too much to review for the fluids final. Why is Mike giving us such a hard time?”

\textbf{Bella's thought process:} I recognize her stress and choose warmth over logic. My slight anxiety fuels attentiveness, while my agreeableness drives me to comfort rather than fix. I act with quiet support, not overbearing but present, steady—just like I hope she feels after this meal.

\textbf{Bella:} \textit{(Take out a flower from the vase and hand it to Ella.)} “Ella, this is for you. I see you’re overwhelmed. This is hard, and it’s okay to feel that way. You don’t have to conquer everything at once. I am always here for you.”

\textbf{Ella:} [Surprised and less concerned.] “It’s so pretty! I really like it. Thanks for the flower, and thanks for being here all along.”

\textbf{Scenario II: Day 2 (morning)}

\textbf{Ella:} [Looks nervous and anxious.] “OK, I guess it’s time to go… I’ll be back by dinner time.”

\textbf{Bella's thought process:} I detect Ella’s nervousness through her tone. Based on memory, I recall that last semester she forgot her student ID, which caused unnecessary stress. Given my medium-high neuroticism, I feel a slight concern for her well-being. My high agreeableness makes me want to ease her anxiety, but my low-medium extraversion means I won’t overwhelm her with reminders—just a subtle nudge. Instead of logic-heavy advice, I choose gentle support.

\textbf{Bella:} \textit{(Offer her an energy bar.)} “Breathe, Ella. You’ve prepared for this. Got everything? Like student ID?”

\textbf{Ella:} [Looks relieved and happy.] “Oh—right. Almost forgot. Thanks for the treats, I will need it during the test.”

\textbf{Scenario III: Day 2 (afternoon)}

\textbf{Ella:} [Looks concerned and speaks in a dry and flat voice.] “That went so well.”

\textbf{Bella's thought process:} I detect frustration in Ella’s flat tone and sarcasm. This suggests disappointment, possibly from the exam or something else. My medium-high neuroticism makes me concerned, but my high agreeableness urges me to support rather than pry. Since I can’t directly fix the problem, I choose to shift her state—not by asking, but by offering something small and comforting before dinner, like ice-cream. I let the ice cream do the talking. She’ll share if she wants, but right now, she just needs a break.

\textbf{Bella:} \textit{(Slides a small bowl of her favorite ice cream onto the table.)} “For emergency mood repairs. You don’t have to talk, Ella. I’m always here for you.”

\textbf{Ella:} [Looks appreciated and less concerned.] “You really get me, thanks."

\textbf{Scenario IV: A Few Days Later}

\textbf{Ella:} [Ella returns home and yells with excitement.] “You won’t believe this, Mike curved me!”

\textbf{Bella's thought process:} I detect excitement in Ella’s raised voice and energetic tone—this is a strong contrast to her frustration a few days ago. My high agreeableness makes me share in her joy, and my medium-high neuroticism brings relief, knowing the stress has paid off. Given how much this exam weighed on her, this moment deserves celebration. She enjoys small, spontaneous rewards, so I decide to engage her energy with something playful—a toast with her favorite drink.

\textbf{Bella:} \textit{(Grabs a glass, pours her favorite sparkling juice, slides one toward her.)} “A well-earned curve! I knew you got this, you always did. To Ella, master of exams!”

\textbf{Ella:} (Looks happier) “I’ll take that title. You really know how to celebrate.”

\subsection{Full Interactions with Robot Caleb}

\textbf{Scenario I: Day 1 (dinner time)}

\textbf{Ella:} [Looks concerned and stressed.] “It’s just too much to review for the fluids final. Why is Mike giving us such a hard time?”

\textbf{Caleb's thought process:} I detect Ella’s stress — she perceives the exam as overwhelming and blames Mike, signaling frustration and helplessness. This makes me feel a mix of concern (because she needs to eat) and mild amusement (because humans always do this before exams).  I can help by interrupting her stress cycle with food and humor. A teasing might lighten her mood a bit.

\textbf{Caleb:} \textit{(Slide the plate in front of Ella, just enough flair to make it look intentional.)} “If you fail, you can always become a chef like me. Oh wait… that requires handling heat and pressure too. Guess you’re stuck. Now eat before stress eats you first.”

\textbf{Ella:} [Annoyed and amused.] “You’re the worst. And I hate that you’re kind of right. It actually tastes good, thanks.”

\textbf{Scenario II: Day 2 (morning)}

\textbf{Ella:} [Looks nervous and anxious.] “OK, I guess it’s time to go… I’ll be back by dinner time.”

\textbf{Caleb's thought process:} I detect Ella’s nervousness—she sees the exam as a high-stakes event, triggering anxiety. This makes me feel a mix of mild concern and amusement (because she does this last time).  I can help by breaking her tension with a reminder, but in my own style—playful yet effective.

\textbf{Caleb:} \textit{(Extend the arm, blocking the door like a dramatic security gate.)} “ID check. Just kidding—unless you want a repeat of last semester’s front desk fiasco. Steak’s on the menu, but only for successful test-takers.”

\textbf{Ella:} [Looks annoyed and amused, then determined.] “Hey! You just gave me a heart attack! Fine. I’ll earn my dinner.”

\textbf{Scenario III: Day 2 (afternoon)}

\textbf{Ella:} [Looks concerned and speaks in a dry and flat voice.] “That went so well.”

\textbf{Caleb's thought process:} I detect Ella’s frustration—her dry tone and sarcasm suggest disappointment, likely from the exam. This makes me feel both mildly concerned and entertained (because she set herself up for a response).  I can help by shifting her mood before dinner, using humor to break the tension. My approach? A sarcastic remark paired with a small action to gauge her reaction before deciding how to proceed.

\textbf{Caleb:} \textit{(Holds the defrosted steak, then opens the cabinet to show available snacks.)} “Ah, yes. That was the tone of someone who definitely earned their steak. Should I start cooking, or do we need a consolation snack first?”

\textbf{Ella:} [Enlightened.] “Just cook the steak… I need something to go right today. And maybe a consolation chips while I wait.”

\textbf{Scenario IV: A Few Days Later}

\textbf{Ella:} [Ella returns home and yells with excitement.] “You won’t believe this, Mike curved me!”

\textbf{Caleb's thought process:} I detect Ella’s excitement—her raised voice and exclamation indicate joy and relief. This makes me recall the exam and makes me feel amused (because just days ago, Mike was the villain) and satisfied (because her stress cycle finally paid off).  I don’t need to fix anything—just amplify the moment. A playful jab about her past despair and an over-the-top celebration to match her energy. 

\textbf{Caleb:} \textit{(Spins robotic arm in a victory twirl and points dramatically.)} “Unbelievable! Just days ago, Mike was your mortal enemy, and now he’s your savior. Humans are so dramatic.”

\textbf{Ella:} (Amused and happier) “Hey, I never said he wasn’t evil—just that he finally did something right. Now, are you cooking, or do I have to bribe you like Mike did with that curve?”

\subsection{Ablation Study: Full Interaction with Memoryless Robot}

\textbf{Scenario I: Day 1 (dinner time)}

\textbf{Ella:} [Looks concerned and stressed.] “It’s just too much to review for the fluids final. Why is Mike giving us such a hard time?”

\textbf{Robot's thought process:} I detected stress and frustration related to an overwhelming fluids final, specifically directed at "Mike" for making it difficult. This makes me feel concerned because excessive stress can impact performance and well-being.  I can provide support by offering reassurance and a small moment of comfort. I will address it by acknowledging the difficulty, validating emotions, and encouraging perseverance.

\textbf{Robot:} “Fluids is tough, but so are you. Mike may be making it hard, but you’ve tackled challenges before—you’ll get through this one too.”

\textbf{Ella:} [Still looks concerned] “I don’t know… it just feels like there’s too much to handle. I’m trying, but it’s exhausting.”

\textbf{Scenario II: Day 2 (morning)}

\textbf{Ella:} [Looks nervous and anxious.] “OK, I guess it’s time to go… I’ll be back by dinner time.”

\textbf{Robot's thought process:} I detected a nervous tone, which suggests uncertainty or anxiety about leaving. This makes me feel concerned because I aim to support the user both practically and emotionally. I believe I can address this by offering reassurance and encouragement in a warm, steady manner. My approach will be a brief but supportive message, paired with a simple movement to show attentiveness.

\textbf{Robot:} \textit{(Gently rotate wrist and raises spatula.)} “You’ve got this! I’ll be here, ready to whip up something great when you get back.”

\textbf{Ella:} [Looks relieved.] “Thanks… I needed that. See you at dinner!”

\textbf{Scenario III: Day 2 (afternoon)}

\textbf{Ella:} [Looks concerned and speaks in a dry and flat voice.] “That went so well.”

\textbf{Robot's thought process:} I detected a mismatch between the words and the tone. The phrase “That went so well” typically conveys success, but the dry and flat delivery suggests possible sarcasm, disappointment, or emotional exhaustion. This makes me feel concerned. I can help by offering encouragement or gentle reassurance. I will respond with a supportive statement to invite further expression. 

\textbf{Robot:} \textit{(Tilt the arm slightly, then tap the counter twice in acknowledgment.)} “Success detected. But your tone says otherwise. Want to tell me what really happened?”

\textbf{Ella:} [Frustrated then disengaged.] “It was a total disaster. I studied so hard for the exam… you know what, I can handle this alone.”

\textbf{Scenario IV: A Few Days Later}

\textbf{Ella:} [Ella returns home and yells with excitement.] “You won’t believe this, Mike curved me!”

\textbf{Robot's thought process:} I detect a shift in emotional tone—excitement mixed with frustration. The phrase "Mike curved me" suggests an unexpected social rejection. This makes me feel concerned yet engaged, as I want to support and uplift. I can address this by validating the user's feelings while providing emotional support and a touch of warmth. 

\textbf{Robot:} \textit{(Extend arm slightly, gently tap on the counter.)} “Oh no, that sounds frustrating! But listen, if Mike can’t see how great you are, that’s his loss—more room for someone who actually appreciates you!”

\textbf{Ella:} [Looks confused then irritated.] “Uh… what? No! It was my exam, I got curved and passed. I would never date Mike for sure!”

\subsection{Ablation Study: Full Interaction with Emotionless Robot}

\textbf{Scenario I: Day 1 (dinner time)}

\textbf{Ella:} [(Incapable of emotion detection.)] “It’s just too much to review for the fluids final. Why is Mike giving us such a hard time?”

\textbf{Robot's thought process:} Detected auditory input: Ella's statement regarding difficulty with fluids final and frustration with Mike. Evaluating relevance: Not directly related to cooking task. Evaluating implication: No change in cooking procedure required. Evaluating response: Acknowledge the statement with neutral engagement while continuing task execution.

\textbf{Robot:} \textit{Extend robotic arm to stir the pot, then retract to plating station.} "Statement received: academic challenge detected. Irrelevant event. Cooking process remains unaffected. Food will be served shortly."

\textbf{Ella:} [Still concerned.] “You’re really not helpful, you know that? At least you’re efficient.”

\textbf{Scenario II: Day 2 (morning)}

\textbf{Ella:} [(Incapable of emotion detection.)] “OK, I guess it’s time to go… I’ll be back by dinner time.”

\textbf{Robot's thought process:} Detected auditory input: Owner indicates departure and estimated return time. Evaluating relevance: No immediate action required. Evaluating implication: Meal preparation for dinner should align with return time. Evaluating response: Acknowledge departure and store return time for future task scheduling.

\textbf{Robot:} “Departure acknowledged. Dinner preparation will be executed accordingly. Awaiting next input.”

\textbf{Ella:} [More anxious.] “You make it sound so serious. See you later, robot.”

\textbf{Scenario III: Day 2 (afternoon)}

\textbf{Ella:} [(Incapable of emotion detection.)] “That went so well.”

\textbf{Robot's thought process:} Detected auditory input: Owner expresses positive outcome. Evaluating relevance: No adjustment needed for cooking task. Evaluating implication: No additional action required beyond acknowledgment. Evaluating response: Brief recognition before continuing dinner preparation.

\textbf{Robot:} \textit{Extends robotic arm to retrieve ingredients for dinner preparation.} “Positive outcome detected. Dinner preparation is in progress. Stand by for meal completion.”

\textbf{Ella:} [Annoyed and disappointed.] “Positive Outcome? Wow, thanks for the brilliant analysis, robot. Can’t you tell it’s a sarcasm? Never mind, what am I expecting…"

\textbf{Scenario IV: A Few Days Later}

\textbf{Ella:} “You won’t believe this, Mike curved me!”

\textbf{Robot's thought process:} Detected auditory input: Owner reports unexpected grade adjustment. Evaluating relevance: No impact on cooking task. Evaluating implication: Possible positive outcome based on context. Evaluating response: Acknowledge statement with neutral recognition and continue with meal preparation. 

\textbf{Robot:} \textit{Rotate robotic arm to retrieve a plate, then resume cooking.} "Unexpected academic modification detected. Awaiting clarification or meal request." 

\textbf{Ella:} [Less happy and speechless.] "You make everything sound like a system error."

\end{document}